%% file: plb_pi0.tex
\documentclass[final]{elsart}

\usepackage{graphicx}

\usepackage{amssymb}

\newcommand{\numu}{\nu_{\mu}}

\newcommand{\nue}{\nu_e}

\begin{document}
\begin{frontmatter}



\title{First Observation of Coherent $\pi^0$ \\ Production in Neutrino Nucleus Interactions with $E_{\nu}<$ 2~GeV}


\input{authors.tex}

\begin{abstract}
The MiniBooNE experiment at Fermilab has amassed the largest sample to date of $\pi^0$s produced in neutral current (NC) neutrino-nucleus interactions at low energy.  This paper reports a measurement of the momentum distribution of $\pi^0$s produced in mineral oil (CH$_2$) and the first observation of coherent $\pi^0$ production below 2~GeV.  In the forward direction, the yield of events observed above the expectation for resonant production is attributed primarily to coherent production off carbon, but may also include a small contribution from diffractive production on hydrogen.  Integrated over the MiniBooNE neutrino flux, the sum of the NC coherent and diffractive modes is found to be (19.5 $\pm$1.1 (stat) $\pm$2.5 (sys))\% of all exclusive NC $\pi^0$ production at MiniBooNE.  These measurements are of immediate utility because they quantify an important background to MiniBooNE's search for $\nu_{\mu} \to \nu_e$ oscillations.
\end{abstract}

\begin{keyword}

\PACS{14.60.Lm, 12.15.Mm, 13.15+g, 25.30.Pt}
\end{keyword}
\end{frontmatter}

\section{Introduction}
Neutral current (NC) $\pi^0$ production is the single largest $\numu$-induced background to neutrino experiments measuring $\numu \rightarrow \nue$ oscillations in the $E_{\nu}\sim 1$~GeV range, including the search recently performed by the MiniBooNE experiment~\cite{AguilarArevalo:2007it}\@.  NC $\pi^0$ events can mimic $\nue$ signal events when, for example, one of the two photons associated with the $\pi^0 \rightarrow \gamma \gamma$ decay is not detected.  This can happen when a photon exits the detector before showering or does not have enough energy to initiate a shower.  Estimating the rate of such backgrounds relies on knowledge of neutrino induced NC $\pi^0$ production at low energy ($E_\nu < 2$~GeV).

Pion production from the scattering of low energy neutrinos on nuclei principally occurs through two mechanisms.  The larger contribution comes from incoherent processes where the neutrino interacts with one of the nucleons in the nucleus.  In the MiniBooNE energy range this mainly consists of the excitation and subsequent pionic decay of baryonic resonances (such as the $\Delta(1232)$).  Additionally there is a small but non-negligible yield from coherent scattering where the neutrino interacts with the entire nucleus leaving it in its ground state.  Because of the necessarily small momentum transfer, coherent $\pi^0$ events are more forward peaked than their resonantly produced counterparts.  

To predict the full spectrum of $\pi^0$ production reliably, it is important to characterize the resonant $\pi^0$ contribution, as it is the dominant source of $\pi^0$s.  MiniBooNE models resonantly produced NC $\pi^0$ events using the Rein and Sehgal (RS) model~\cite{Rein:1980wg} as implemented in version 3 of the \textsc{Nuance} event generator~\cite{Casper:2002sd} assuming an $N\rightarrow\Delta$ dipole form factor with axial mass, $M_A^{res}=1.1$~GeV/c$^2$.  For MiniBooNE, $95\%$ of resonant NC $\pi^0$ production is predicted to occur via the $\Delta(1232)$, but seventeen higher mass resonances with their interferences also contribute in the model.  $85\%$ of the resonant NC $\pi^0$ production at MiniBooNE should occur on carbon with the remaining 15\% on hydrogen.  To predict both the kinematics and yield of coherently produced $\pi^0$ events, MiniBooNE uses the RS coherent production model~\cite{Rein:1982pf}, implemented in \textsc{Nuance} with the relevant axial mass set to $M_A^{coh}=1.03$~GeV/c$^2$.  The model predicts coherent $\pi^0$ production to be 30\% of the total NC exclusive $\pi^0$ production in MiniBooNE.  The \textsc{Nuance} implementation differs from the RS model in two important ways.  First, resonances are decayed isotropically, which is not strictly correct.  Events are reweighted to match the RS model based on the $\Delta$ decay angle in its rest frame with respect to its momentum vector.  Second, Rein and Sehgal describe an absorption factor, which scales the coherent production cross section for NC $\pi^0$s, while in \textsc{Nuance}, absorption is implemented as part of the final state interaction (FSI) model.  The designations of resonant and coherent are set prior to any FSI\@, which means that rescattered events with a $\pi^0$ in the final state may be misclassified in \textsc{Nuance}, as would be the case when a coherently produced $\pi^0$ rescatters elastically through a resonance.

Calculating $\pi^0$ production cross section in either case becomes complicated for several reasons.  In the case of resonant $\pi^0$ production, the neutrino-nucleon cross section requires knowledge of the appropriate transition form factors.  Using the CVC hypothesis~\cite{Feynman:1958,Gerstein:1956}, vector form factors can be reliably inferred from electron scattering data; however, axial-vector form factors are not well known and rely heavily on the use of PCAC~\cite{Goldberger:1958tr}\@.  Added to this, for neutrino-nucleus scattering, one must additionally calculate the propagation of $\pi^0$'s through the target nucleus as this can change both the identity of the pion as well as its kinematics.  At resonance energies, the pion-nucleon cross section is large and hence the pion produced in the resonance decay has a non-negligible probability to re-interact before exiting the target nucleus.  This must be properly accounted for in any useful simulation, because the experimental observable is the $\pi^0$ only after it has exited the target nucleus.  

In the case of coherent scattering, the situation is even more ambiguous.  Calculations of coherent scattering cross sections have been performed using detailed neutrino-nucleon resonance production models and subsequent hadron-hadron interactions~\cite{Kelkar:1996iv,AlvarezRuso:2007tt}\@.  Such a recent calculation~\cite{AlvarezRuso:2007sq} finds a value of 14\% for the ratio of coherent to incoherent scattering for the NC process investigated here.  Alternatively, one can circumvent some of the complexities of these dynamics by invoking Adler's PCAC theorem~\cite{Adler:1964yx}, which directly identifies the coherent scattering cross section with the elastic scattering of pions on the same nucleus.  This procedure~\cite{Rein:1982pf,Kelkar:1996iv,Belkov:1986hn,Marteau:1999zp,Paschos:2005km,Singh:2006bm} works well at higher energies, but appears to fail at low energies where PCAC-based calculations typically predict a substantially larger fraction of coherent scattering.  For example, using data on $\pi+C$ scattering~\cite{Ashery:1981tq}, one would infer a coherent fraction of roughly 50\%\@.  In addition to the differing theoretical approaches and large range in coherent scattering predictions, the K2K experiment has recently, and somewhat surprisingly, reported no evidence for charged current coherent $\pi^+$ production at 1.3~GeV~\cite{Hasegawa:2005td}\@.  For these reasons, experimental measurements of coherent pion production are critical to our understanding of this complex process, and are especially important at low energy.

To date, there are only a few measurements of NC $\pi^0$ production in the 1-2 GeV energy range, conducted on a variety of different targets materials, which together consist of about $3000$ events~\cite{Hawker:2005vf,Krenz:1977sw,Faissner:1983ng,Nakayama:2004dp}.  NC coherent $\pi^0$ production data are even more sparse.  Although cross section measurements exist at higher energies~\cite{Vilain:1993sf}, there are no measurements of NC coherent $\pi^0$ production below $E_\nu<2$ GeV, which is an important region for neutrino oscillation experiments.  

The remainder of this paper describes the MiniBooNE experimental setup, the identification and reconstruction of NC $\pi^0$ events, a measurement of the overall yield of NC $\pi^0$ production in mineral oil as a function of $\pi^0$ momentum, and a direct measurement of the coherent $\pi^0$ production fraction in this data sample. Understanding coherent production is critical in reproducing the observed angular spectrum of these events. Together, this technique and resultant $\pi^0$ constraints provide important input to the MiniBooNE oscillation analysis~\cite{AguilarArevalo:2007it} and update previously reported work on this subject~\cite{Raaf:2004ty,Raaf:2005up,Link:2007mf}.

\section{The Experiment}
The MiniBooNE $\nu_{\mu}$ beam results from the decays of secondary particles (mostly pions) that are produced by interactions of 8~GeV protons from Fermilab's Booster  incident on a beryllium target.  The detector, 541~m downstream of the beryllium target, is a 12.2~m diameter spherical tank filled with 800 tons of pure mineral oil.   The tank is separated into two regions: an inner volume with a radius of 575~cm, and a 35~cm thick outer volume.  An optical barrier provides the separation of the two regions, and also serves as the support structure for 1280 equally-spaced 8-inch photomultiplier tubes (PMTs) that give 10\% photocathode coverage of the inner volume.  PMT hits have a threshold of $\sim$0.1 photoelectrons, and are recorded in a 19.2~$\mu$s window around every 1.6~$\mu$s neutrino beam spill.  An additional 240 8-inch PMTs mounted in the outer volume act as a veto shield to detect charged particles entering or exiting the detector.  Three meters of dirt above the detector give a 60\% reduction in cosmic ray flux, and with appropriate selection cuts, the outer veto region rejects more than 99.95\% of the cosmic rays observed in the detector. 

The $\nu_{\mu}$ energy spectrum peaks at 700~MeV and extends to approximately 3~GeV.  Integrated over the neutrino flux, approximately $7\%$ of the neutrino interactions in MiniBooNE are predicted to be NC exclusive $\pi^0$ production.

\section{Event Selection and Analysis}
The selection of NC $\pi^0$ events begins with a set of simple pre-reconstruction cuts (or pre-cuts) that exactly match those used in the MiniBooNE $\nu_e$ appearance analysis~\cite{AguilarArevalo:2007it}.  Each candidate must have only a primary event, with no evidence of a secondary event consistent with a muon decay electron.  This eliminates the vast majority of CC $\nu_{\mu}$ interactions.  Each event is also required to have more than 200 PMT hits in the main tank, well above the maximum number of hits observed at the muon-decay endpoint.  Each event must have fewer than 6 hits in the veto region, which eliminates cosmic rays and neutrino events not contained in, or originating outside of, the detector. Additionally, all events must be in the 1.6~$\mu$s beam spill window, although, after the preceding cuts are applied, almost no events exist outside the beam window.

Each event that passes the pre-cuts is then reconstructed under three hypotheses~\cite{Patterson:2007zz}: muon, electron, and $\pi^0$.  The reconstruction is based on the expected distribution of \v{C}erenkov and scintillation light in the mineral oil.  Under the muon and electron hypotheses, the times and charges of hit PMTs are used to reconstruct a single track fitting the location and time of the neutrino interaction as well as the energy and direction of the charged lepton track.  Electron and muon tracks are distinguished by the energy deposition per unit length and by the sharpness of the \v{C}erenkov ring edge (electron rings are fuzzier than muon rings due to scattering and the formation of  electromagnetic showers). The $\pi^0$ hypothesis requires a two track fit which fits the location and time of the neutrino interaction as well as the energies, directions, and conversion distances of the two photons.  Each photon track fit assumes that the light is distributed in the manner of an electron track.  Each fitted hypothesis produces a likelihood ($\mathcal{L}_{\mu}$, $\mathcal{L}_e$, $\mathcal{L}_{\pi}$), and the logs of the ratio of likelihoods from different hypotheses are used for particle identification.  The $\pi^0$ fit is run in two ways: with a floating invariant mass, $m_{\gamma\gamma}$, and with $m_{\gamma\gamma}$ fixed to the $\pi^0$ mass.  The $\pi^0$ parameters, with the exception of the invariant mass, are obtained from the fixed mass fit as this provides the most accurate estimate of the true $\pi^0$ kinematics.

The reconstructed parameters and likelihoods allow for further selection.  Because their decay photons shower like electrons, $\pi^0$ events should look more like electrons than muons, and overall these events should look more like a $\pi^0$ than an electron.  Therefore, particle identification cuts requiring $\log(\mathcal{L}_e/\mathcal{L}_{\mu}) > 0.05 $\@ and $\log(\mathcal{L}_{\pi}/\mathcal{L}_e) > 0$ are applied.  Additionally, a fiducial volume cut requires the reconstructed position of the event to be within 500~cm of the detector center.  These cuts produce a very clean sample of $\pi^0$ events with a signal to background ratio of $\sim$30.  With an additional selection on the invariant mass ($80 < m_{\gamma\gamma} < 200$~MeV/c$^2$), the $\pi^0$ efficiency predicted by the Monte Carlo (MC) is 39.5\%, as detailed in Table~\ref{survival}.  After all cuts, the MiniBooNE data set consists of 28,000 NC $\pi^0$ events produced in 5.6$\times$10$^{20}$ protons on target, which is the largest sample of NC $\pi^0$ events yet collected at these energies.
\begin{table}[tb]
\centerline{\begin{tabular}{@{\extracolsep{2.0cm}}lr}
\hline
Cut Level  &
\multicolumn{1}{c}{Survival} \\
\hline
Pre-cuts                                         &  100\% \\
$r<$500~cm cut                                   & 79.4\% \\
$\log(\mathcal{L}_e/\mathcal{L}_{\mu}) > 0.05$   & 51.5\% \\
$\log(\mathcal{L}_{\pi}/\mathcal{L}_e) > 0$      & 50.1\% \\
80$< m_{\gamma\gamma} <$200~MeV/c$^2$            & 39.5\% \\
\hline
\end{tabular}}
\caption{\label{survival} NC $\pi^0$ event survival fraction (in percent) for each selection requirement relative to the pre-cuts.  The cuts are applied progressively from the top of the table down.}
\end{table}

Once selected, the $\pi^0$ candidate events are divided into bins of reconstructed $\pi^0$ momentum and the MC is used to unsmear the data ({\it i.e.}, to reverse the effects of momentum resolution and inefficiency and thus obtain the ``true'' $\pi^0$ production rates as a function of momentum).  A matrix is formed by dividing MC events into bins of true momentum versus reconstructed momentum and counting true $\pi^0$ events in each bin.  Events in this matrix must pass all selection cuts including the mass window cut.  A MC event is defined to be a $\pi^0$ event if at least one decayed $\pi^0$ exists in the final state.  This definition includes both NC and CC events, although most CC events are eliminated by the pre-cuts which exclude events with electrons from muon-decay.  The event count in each bin is divided by the total number of $\pi^0$ events in that true momentum bin (including events that did not pass the reconstruction cuts, which are included in the denominator to correct for the cut efficiency).  This matrix is inverted to form the unsmearing matrix.  Next, a vector is formed in the data by separating the $\pi^0$ candidate events into the same reconstructed bins.  The background rate in each data bin is estimated using the MC and the data event yields are scaled to remove this estimated background.  The product of the unsmearing matrix and the data vector is the unsmeared data vector.  While in many applications this kind of matrix unsmearing can be unstable, leading to large uncertainties in the unsmeared quantities~\cite{cowan}, in this case, with the initial matrix largely diagonal, the process is quite reliable.  This assertion is supported both by MC closure tests and by the reasonable size of the errors on the unsmeared data as propagated through the matrix.  

\begin{figure*}[tb!]
\centerline{\includegraphics[width=13.5cm]{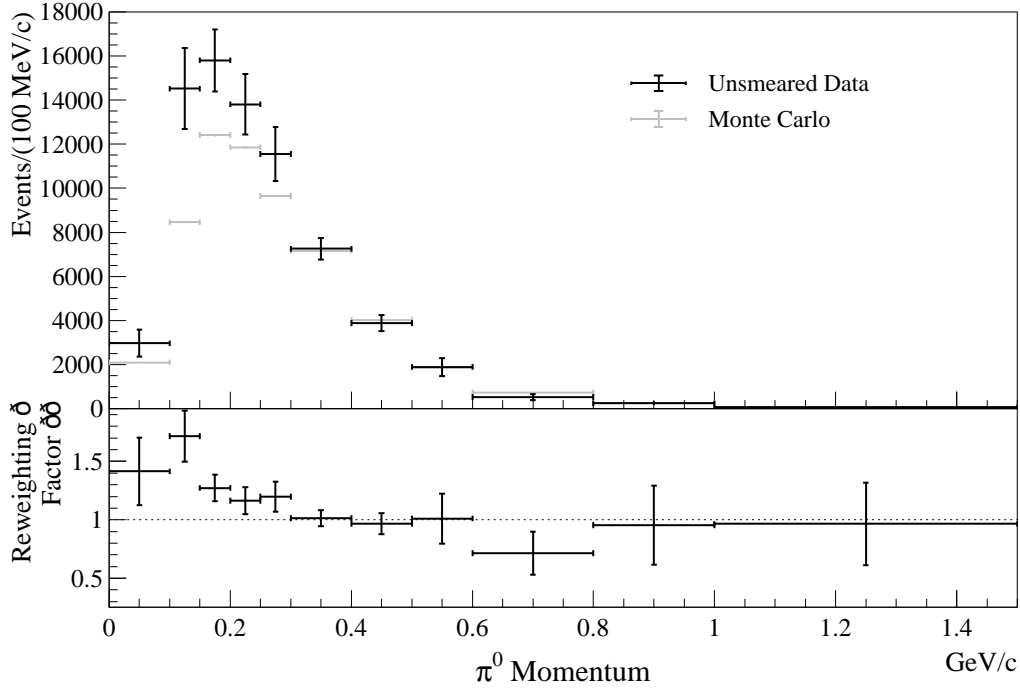}}
\caption{\label{mom_corr1}Top: Results of the $\pi^0$ unsmearing in bins of momentum.  The dark points show the unsmeared data $\pi^0$ momentum distribution and the light points show the uncorrected MC $\pi^0$ momentum distribution.  The unsmeared data error bars contain all sources of error propagated through the unsmearing, while the MC error bars result solely from finite MC statistics.  Bottom: The reweighting function, formed by taking the ratio of the two distributions in the top plot (data/MC).}
\end{figure*}

Figure~\ref{mom_corr1} shows a comparison of the initial MC prediction to this unsmeared data distribution.  The ratio of the two distributions forms a reweighting function which is used to scale MC $\pi^0$ events as a function of true momentum.  This distribution reflects the extent to which the starting MC does or does not predict the measured momentum spectrum of $\pi^0$ events in MiniBooNE.

\begin{figure*}[tb!]
\centerline{\includegraphics[width=13.5cm]{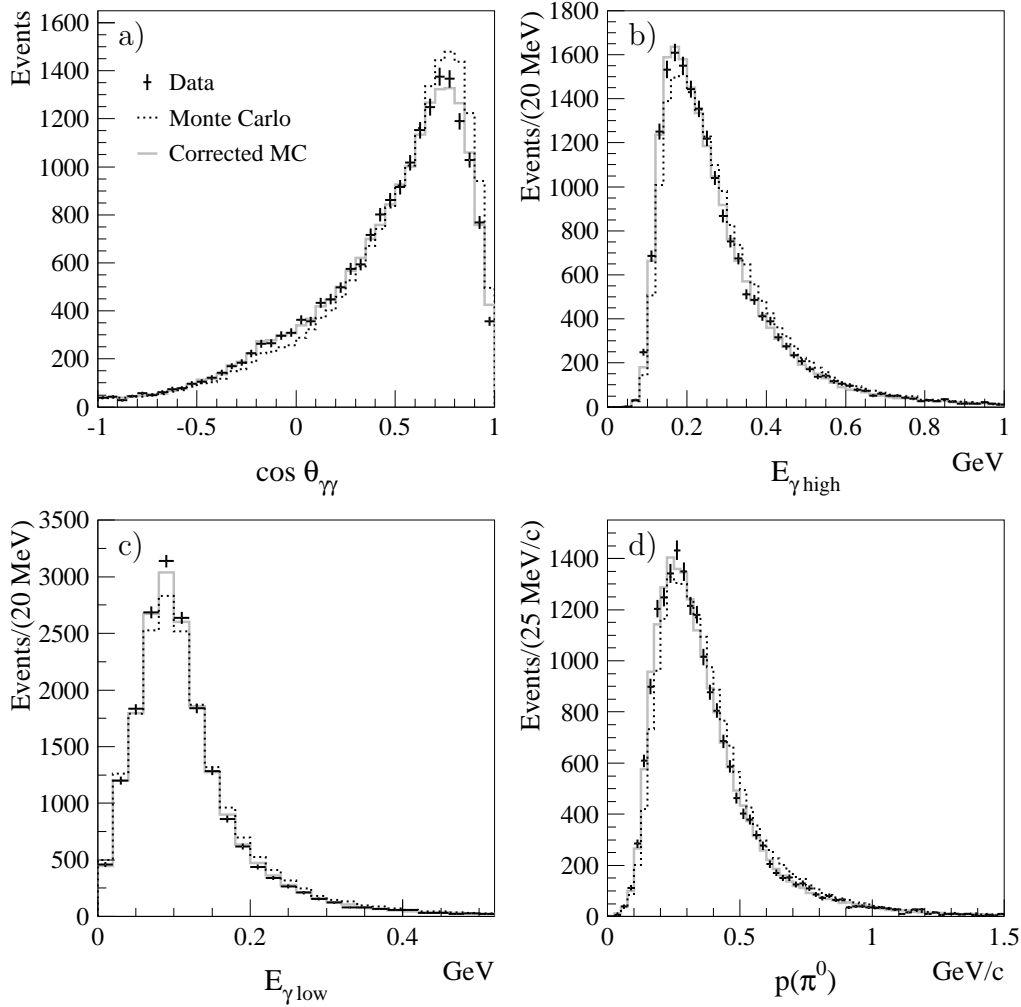}
\begin{picture}(0,0)
\put(-345,365){a)}
\put(-152,365){b)}
\put(-345,172){c)}
\put(-152,172){d)}
\end{picture}}
\caption{\label{mom_corr3} Relatively normalized comparison of the uncorrected (dotted) and momentum corrected (solid gray) Monte Carlo to data (points with statistical error bars) for various $\pi^0$ kinematic distributions: a) the opening angle between the two photons in the $\pi^0$ decay, b) energy of the more energetic photon, c) energy of the less energetic photon, and d) the $\pi^0$ momentum.  The marked improvement in these kinematic distributions shows that the initial data-to-MC differences can largely be attributed to the original disagreement in $\pi^0$ production as a function of momentum.}
\end{figure*}

By construction, this momentum reweighting fixes the discrepancy between data and MC in reconstructed $\pi^0$ momentum.  Additionally, it improves agreement in many key kinematic distributions.  Figure~\ref{mom_corr3} shows relatively normalized data to MC comparisons for both the initial and corrected MC.  The kinematic distributions shown are the cosine of the $\gamma\gamma$ opening angle, the photon energies, and the $\pi^0$ momentum.  All distributions show marked improvement (this is almost a tautology for the momentum distribution, except that it is binned more finely than the correction function and is in terms of reconstructed momentum).  This reweighting of the MC is neither profound nor forbidden; it merely addresses a range of imperfections in the simulation, which may include contributions from the neutrino flux to the $\pi^0$ production model.  

\section{Results}
The $\pi^0$ candidate events in the momentum reweighted MC are divided into three templates: one each for resonant ($res$), coherent ($coh$), and background ($bg$) events.  The resonant and coherent templates contain all true exclusive $\pi^0$ events\footnote{By our definition, a true exclusive $\pi^0$ event must be generated by \textsc{Nuance} in either the coherent or resonant modes and there must be a decayed $\pi^0$ in the final state.}.  In the case of the coherent template, this includes diffractive scattering off hydrogen~\cite{Rein:1986cd}, which for MiniBooNE is predicted to be 16\% of the coherent template.  The background template consists of all other events, including some events that contain one or more decayed $\pi^0$ not produced in the resonant or coherent channels.

The templates are formed as a two dimensional (2D) distribution of
$E_{\pi}(1-\cos\theta_{\pi})$ versus invariant mass $m_{\gamma\gamma}$\@.  Use of this 2D distribution helps to break the degeneracy between the coherent and resonant templates in $m_{\gamma\gamma}$, and between the coherent and background templates in $E_{\pi}(1-\cos\theta_{\pi})$\@.  The angle $\theta_{\pi}$ is defined to be the lab angle of the reconstructed $\pi^0$ momentum vector with respect to the neutrino beam direction\footnote{The simpler angular function, $\cos\theta_{\pi}$, was tried in place of $E_{\pi}(1-\cos\theta_{\pi})$ and found to have slightly poorer performance.  This is attributed to the fact that $E_{\pi}(1-\cos\theta_{\pi})$ has a more consistent shape across all $\pi^0$ momenta for coherent events.}.  The 2D template binning is defined by dividing the 1D distributions ($E_{\pi}(1-\cos\theta_{\pi})$ and $m_{\gamma\gamma}$) from the MC into variable-width bins of approximately equal numbers of events.  Each fit has three parameters ($x^{res}$, $x^{coh}$ and $x^{bg}$) which scale the template distributions independently.  The fit minimizes the following $\chi^2$:
\begin{equation}
\chi^2 = \sum_i \frac{[f_i^{data}-\!(f_i^{res}x^{res}\!+\!f_i^{coh}x^{coh}\!+\! f_i^{bg}x^{bg})]^2}{(\sigma_i^{data})^2 + (\sigma_i^{res})^2 + (\sigma_i^{coh})^2 + (\sigma_i^{bg})^2}
\end{equation}
where $f_i^{\alpha}$ is the fraction of total events of type $\alpha$ (where $\alpha$ is $data$, $res$, $coh$, and $bg$) in the $i$th bin and $\sigma_i^{\alpha}$ is the statistical uncertainty on that fraction.  Since the fit is to the fractional distributions, it is a shape only fit and the sum of the fit parameters should be very close to unity, as is the case for all fits.

\begin{figure*}[tb!]
\centerline{\includegraphics[width=13.5cm]{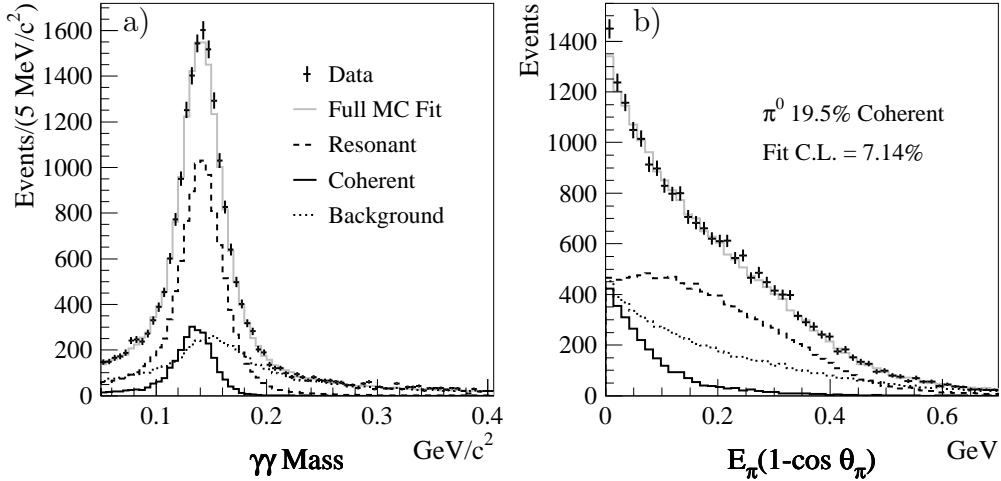}
\begin{picture}(0,0)
\put(-341,173){a)}
\put(-148,173){b)}
\end{picture}}
\caption{\label{coh_fit} Comparison of data and Monte Carlo in a) $m_{\gamma\gamma}$ invariant mass, and b) $E_\pi(1-\cos\theta_{\pi})$ after the coherent fraction fit.  The resonant, coherent and background components are shown scaled by their fit parameters.  The full MC fit is the sum of the three components.}
\end{figure*}

The fit is repeated for several different binnings.  The number of bins in each 1D projection is varied independently from 15 to 25, for a total of 121 different binning combinations.  The final fit parameters are formed from the average of the parameters from the 121 fits.  To determine the best overall production parametrization, the momentum correction and coherent fit are iterated, using the results of one as a correction to the inputs of the other.  This procedure converges after only two iterations.  Figure~\ref{coh_fit} shows the final fit plotted in the $m_{\gamma\gamma}$ and $E_{\pi}(1-\cos\theta_{\pi})$ projections.  The fit coherent fraction is defined as:
\begin{equation}
F_{coh} = \frac{x^{coh}}{x^{coh} + x^{res}} \times 100\%.
\end{equation} 

The fit finds $F_{coh}$ to be $(19.5 \pm 1.1 (stat)\%$\@.  The MiniBooNE data clearly favor the presence of a coherent scattering component.  The average confidence level (C.L.) of the fit is $7.14\%$,while the C.L. obtained when the coherent fraction is fixed to zero $(x^{coh} \equiv 0)$ is $10^{-18}$.  The effects of the momentum reweighting are small, but not insignificant.  If the momentum reweighting is not done the fit coherent fraction is 18.5\%.  

\begin{figure}[t]
\centerline{\includegraphics[width=7.5cm]{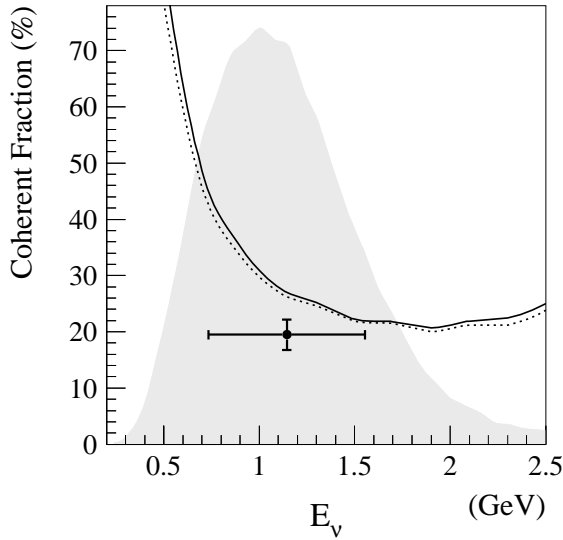}}
\caption{\label{compare_rs}The coherent fraction in the Rein and Seghal based MC vs.\ neutrino energy compared to this measurement.  The solid line includes only carbon interactions, while the dotted line includes scattering off hydrogen with diffractive events counted as part of the coherent.  The measured value is shown with error bars which indicate the total error on the measurement (vertical) and the spread in the participating neutrino energy distribution (horizontal).  The shaded distribution is MC energy for neutrinos which produce NC $\pi^0$ events in MiniBooNE with arbitrary normalization.  The coherent fraction predicted by \textsc{Nuance} integrated over all energies in MiniBooNE is 30\%.}
\end{figure}

One should note that the reported coherent fraction is specific to the MiniBooNE neutrino spectrum and includes scattering off both carbon and hydrogen nuclei in the mineral oil target.  It has also been measured in the context of the RS-based \textsc{Nuance} generator~\cite{Rein:1982pf,Casper:2002sd}, with the aforementioned modification to the $\Delta$ decay angular distribution.  This widely-used model predicts a coherent fraction of 30\% for MiniBooNE. Of course, more recent calculations of coherent production~\cite{AlvarezRuso:2007tt,Belkov:1986hn,Paschos:2005km} do predict a range
of lower coherent fraction values for MiniBooNE.  Figure~\ref{compare_rs} compares the measured coherent fraction to the RS/\textsc{Nuance} prediction as a function of neutrino energy.  The plot shows two predictions: one with both carbon and hydrogen scattering (dashed) and another which includes only carbon interactions (solid).  The effect of hydrogen scattering is small compared to the precision of the measured coherent fraction.  Using the MC to correct to a pure carbon target would yield a measured coherent fraction of $(20.3 \pm 2.8 (stat))\%$.  The shaded distribution shows the predicted neutrino energy spectrum for neutrinos which participate in NC $\pi^0$ production in MiniBooNE.  

\section{Systematic Uncertainties}
Systematic uncertainties on the coherent fraction include choice of binning, background composition, momentum reweighting, neutrino flux, choice of analysis cuts, and detector modeling. The binning systematic is deduced from the RMS on the coherent fraction from the fits to the 121 different binnings.  To determine the background shape uncertainty, background events are divided into several classes and the production cross section of each class is randomly varied according to a Gaussian distribution (with standard deviations from 1 to 40\% depending on the estimate of the uncertainty in each process class).  This is repeated 5000 times and the background shape from each combination of variations is used in the template fit.  The background uncertainty is given by the RMS of these 5000 fits.  The reweighting error is determined by randomly varying the momentum reweighting function according to its errors, paying careful attention to bin-to-bin correlations.  Estimating this uncertainty proceeds as in the background case with 5000 random combinations and fits.  The flux uncertainty results from varying parameters in the beam simulation.  The analysis cuts error is determined by varying the cut point on the reconstructed variables.  Finally, the detector model uncertainty is determined by fitting, as fake data, 70 data-sized MC samples which were simulated with random, but properly correlated variations in several detector response parameters.  Since the 70 samples are statistically independent, the detector model error is: 
\begin{equation}
\sigma_{det.\ model} = \sqrt{RMS^2-\langle\sigma_{fit}\rangle^2}
\end{equation}
where $RMS$ is the root mean square of the 70 fits and $\langle\sigma_{fit}\rangle$ is the average fit error of 1.1\%.

Table~\ref{coh_sys} lists the uncertainties estimated from each source.  The dominant source of systematic error is the detector model, which is largely due to the uncertainty in the reconstructed energy which is strongly correlated with several of the varied parameters.

\begin{table}[b]
\centerline{\begin{tabular}{@{\extracolsep{1.0cm}}lr}
\hline
Error Source  &
\multicolumn{1}{c}{Coherent Fraction (\%)} \\ 
\hline 
Binning          & 0.21\\
Background Shape & 0.64\\
Reweighting      & 0.51\\
Flux             & 0.06\\
Analysis Cuts    & 0.51\\
Detector Model   & 2.34\\
\hline
Total Systematic Error  & 2.54\\
\hline
\end{tabular}}
\caption{\label{coh_sys} Contributions to the systematic uncertainty in the  MiniBooNE measured coherent fraction.}
\end{table}

\section{Conclusions}
Using a high statistics sample of events, MiniBooNE has measured the rate of NC $\pi^0$ production in mineral oil as a function of momentum and extracted a correction to the predicted production rate for this process.  MiniBooNE also reports the observed rate of coherent $\pi^0$ production relative to the total exclusive $\pi^0$ production in the context of the RS model~\cite{Rein:1982pf} as implemented in \textsc{Nuance}\cite{Casper:2002sd}.  The coherent fraction is found to be $(19.5\pm 1.1\mathrm{(stat)} \pm 2.5 \mathrm{(sys)})\%$ for the MiniBooNE flux and target.  This should be compared to the 30\% fraction predicted by the RS-based\textsc{Nuance} model, a value significantly higher than the measurement reported here.  The fit to the MiniBooNE NC $\pi^0$ sample excluded, with high confidence, the possibility of no coherent contribution to $\pi^0$ production at MiniBooNE energies.

In the MiniBooNE $\nu_e$ appearance oscillation analysis~\cite{AguilarArevalo:2007it}, both the $\pi^{0}$ momentum correction and the measured coherent fraction were used to reweight the MC for a more accurate estimation of $\pi^0$ misidentification in MiniBooNE\@.  By correcting the $\pi^0$ production with data as described in this paper, the error on the overall $\pi^0$ production (an input to the oscillation analysis) goes from $\sim$25\% (the quadratic sum of a $\sim$20\% flux error and a $\sim$20\% NC $\pi^0$ cross section error) down to 5\%\@.  This represents an important improvement in the sensitivity to $\nu_e$ appearance.

\section{Acknowledgments}
We wish to acknowledge the support of Fermilab, the Department of Energy, and the NSF in the construction, operation, and data analysis of the MiniBooNE experiment.  We also acknowledge the use of the \textsc{Condor} software in the analysis of the data.

\bibliographystyle{elsart-num}
\bibliography{ncpi0}

\end{document}

%% file: authors.tex
\collaboration{The MiniBooNE Collaboration}
\author[columbia]{A.~A. Aguilar-Arevalo},
\author[yale]{C.~E.~Anderson},
\author[princeton]{A.~O.~Bazarko},
\author[fnal]{S.~J.~Brice},
\author[fnal]{B.~C.~Brown},
\author[columbia]{L.~Bugel},
\author[umich]{J.~Cao},
\author[columbia]{L.~Coney},
\author[columbia]{J.~M.~Conrad},
\author[indiana]{D.~C.~Cox},
\author[yale]{A.~Curioni},
\author[columbia]{Z.~Djurcic},
\author[fnal]{D.~A.~Finley},
\author[yale]{B.~T.~Fleming},
\author[fnal]{R.~Ford},
\author[fnal]{F.~G.~Garcia},
\author[lanl]{G.~T.~Garvey},
\author[lanl,fnal]{C.~Green},
\author[indiana,lanl]{J.~A.~Green},
\author[colorado]{T.~L.~Hart},
\author[lanl,cinci]{E.~Hawker},
\author[lsu]{R.~Imlay},
\author[cinci]{R.~A. ~Johnson},
\author[columbia]{G.~Karagiori},
\author[fnal]{P.~Kasper},
\author[indiana]{T.~Katori},
\author[fnal]{T.~Kobilarcik},
\author[fnal]{I.~Kourbanis},
\author[bucknell]{S.~Koutsoliotas},
\author[princeton]{E.~M.~Laird},
\author[yale]{S.~K.~Linden},
\author[vtech]{J.~M.~Link},
\author[umich]{Y.~Liu},
\author[bama]{Y.~Liu},
\author[lanl]{W.~C.~Louis},
\author[columbia]{K.~B.~M.~Mahn},
\author[fnal]{W.~Marsh},
\author[fnal]{P.~S.~Martin},
\author[lanl]{G.~McGregor},
\author[lsu]{W.~Metcalf},
\author[princeton]{P.~D.~Meyers},
\author[fnal]{F.~Mills},
\author[lanl]{G.~B.~Mills},
\author[columbia]{J.~Monroe},
\author[fnal]{C.~D.~Moore},
\author[colorado]{R.~H.~Nelson},
\author[columbia]{V.~T.~Nguyen},
\author[marys]{P.~Nienaber},
\author[lsu]{J.~A.~Nowak},
\author[lsu]{S.~Ouedraogo},
\author[princeton]{R.~B.~Patterson},
\author[bama]{D.~Perevalov},
\author[indiana]{C.~C.~Polly},
\author[fnal]{E.~Prebys},
\author[cinci]{J.~L.~Raaf},
\author[lanl,florida]{H.~Ray},
\author[umich]{B.~P.~Roe},
\author[fnal]{A.~D.~Russell},
\author[lanl]{V.~Sandberg},
\author[lanl]{R.~Schirato},
\author[columbia]{D.~Schmitz},
\author[columbia]{M.~H.~Shaevitz},
\author[princeton]{F.~C.~Shoemaker},
\author[embry]{D.~Smith},
\author[yale]{M.~Soderberg},
\author[columbia]{M.~Sorel},
\author[fnal]{P.~Spentzouris},
\author[bama]{I.~Stancu},
\author[fnal]{R.~J.~Stefanski},
\author[lsu]{M.~Sung},
\author[princeton]{H.~A.~Tanaka},
\author[indiana]{R.~Tayloe},
\author[colorado]{M.~Tzanov},
\author[lanl]{R.~Van~de~Water},
\author[lsu]{M.~O.~Wascko},
\author[lanl]{D.~H.~White},
\author[colorado]{M.~J.~Wilking},
\author[umich]{H.~J.~Yang},
\author[columbia,lanl]{G.~P.~Zeller},
\author[colorado]{E.~D.~Zimmerman}
\address[bama]{University of Alabama, Tuscaloosa, AL 35487}
\address[bucknell]{Bucknell University, Lewisburg, PA 17837}
\address[cinci]{University of Cincinnati, Cincinnati, OH 45221}
\address[colorado]{University of Colorado, Boulder, CO 80309}
\address[columbia]{Columbia University, New York, NY 10027}
\address[embry]{Embry-Riddle Aeronautical University, Prescott, AZ 86301}
\address[fnal]{Fermi National Accelerator Laboratory, Batavia, IL 60510}
\address[florida]{University of Florida, Gainesville, FL 32611}
\address[indiana]{Indiana University, Bloomington, IN 47405}
\address[lanl]{Los Alamos National Laboratory, Los Alamos, NM 87545}
\address[lsu]{Louisiana State University, Baton Rouge, LA 70803}
\address[umich]{University of Michigan, Ann Arbor, MI 48109}
\address[princeton]{Princeton University, Princeton, NJ 08544}
\address[marys]{Saint Mary's University of Minnesota, Winona, MN 55987}
\address[vtech]{Virginia Polytechnic Institute \& State University, Blacksburg, VA 24061}
\address[yale]{Yale University, New Haven, CT 06520}